\documentclass[aps,prb,twocolumn,amsmath,amssymb,groupedaddress,floatfix,longbibliography]{revtex4-1}
\setcitestyle{square,numbers}

\usepackage{graphicx}  
\usepackage{bm}        
\usepackage{amssymb}   
\usepackage{amsmath}
\usepackage{color}
\usepackage{epstopdf}

\usepackage[plainpages=false,pdfpagelabels,colorlinks=true,linkcolor=red,urlcolor=blue,citecolor=blue,pdftitle={Title},pdfauthor={},pdfdisplaydoctitle=true,pdfduplex=DuplexFlipLongEdge]{hyperref}

\begin{document}
\title{Dimensionality of metallic atomic wires on surfaces}

\author{E. Jeckelmann}
\email[E-mail: ]{eric.jeckelmann@itp.uni-hannover.de}
\affiliation{Leibniz Universit\"{a}t Hannover, Institut f\"{u}r Theoretische Physik, Appelstr.~2, 30167 Hannover, Germany}

\date{\today}

\begin{abstract}
We investigate the low-energy collective charge excitations (plasmons, holons)
in metallic atomic wires deposited on semiconducting substrates.
These systems are described by two-dimensional correlated models 
representing strongly anisotropic lattices or weakly coupled chains. 
Well-established theoretical approaches and results are used to study their properties:
random phase approximation for anisotropic Fermi liquids and 
bosonization for coupled Tomonaga-Luttinger liquids 
 as well as Bethe Ansatz and density-matrix renormalization group
methods for ladder models.
We show that the Fermi and Tomonaga-Luttinger liquid  theories predict the same qualitative behavior
for the dispersion of excitations at long wave lengths.
Moreover, their scaling depends on the choice of the effective electron-electron interaction
but does not characterize the dimensionality of the metallic state.
Our results also suggest that such anisotropic correlated systems can exhibit two-dimensional dispersions 
due to the coupling between wires
but remain quasi-one-dimensional strongly anisotropic conductors or retain typical features of
Tomonaga-Luttinger liquids such as the power-law behaviour of the density of states at the Fermi energy.
Thus it is possible that atomic wire materials such as Au/Ge(100)
exhibit a mixture of features associated with one and two dimensional metals. 
\end{abstract}

\maketitle

\section{Introduction}

Atomic wires on semiconductor substrates are prime candidates to
realize one-dimensional (1D) metals~\cite{Springborg07,Oncel08,Dudy17}.
Within the theory of Tomonaga-Luttinger (TL) liquids~\cite{Giamarchi03,Bruus04,solyom3} 
the low-energy behavior of gapless 1D electronic systems
is determined by collective bosonic charge and spin excitations (called holons and spinons, respectively).
The holon excitations are the counterpart of the plasmon excitations predicted by the Fermi liquid theory.
In practice, it is often unclear whether the two-dimensional (2D) arrays of atomic wires 
are better described as (weakly) coupled 1D systems or (strongly) anisotropic 2D metals.
Consequently, the question occurs whether the Fermi liquid theory is enough to explain 
the low-energy electronic properties of metallic atomic wires or the TLL theory 
is necessary to describe correlation effects.

In particular, gold wires on Ge(100) surfaces seem to realize 1D electronic systems~\cite{Meyer11,Dudy17}
and a signature of the TLL theory (the power-law behavior of the density of state at the Fermi energy)
has been found in the scanning tunnelling spectroscopy and photoemission spectra 
of this material~\cite{Blumenstein11a,Meyer14}.
These findings have been contested, however, because
Au/Ge(100) appears to exhibit an anisotropic 2D metallic dispersion at the Fermi energy,
which seems to rule out 1D electronic states and thus the applicability of the TLL
theory~\cite{Nakatsuji11,Nakatsuji12,Park14,DeJong16}.
The signatures of TLLs have also been observed in 
the photoemission spectra of other atomic wires on surfaces such as Bi/InSb(001)~\cite{Ohtsubo15}
and  Pt/Ge(001)~\cite{Yaji16}.

Moreover, low-dimensional plasmons have been found  in several atomic wire systems,   
In/Si(111)~\cite{Hwang07,Liu08}, Pb/Si(557)~\cite{Block11},
Ag/Si(557)~\cite{Krieg13}, Au/Si(557)~\cite{Nagao06}, Au/Si(553)~\cite{Lichtenstein16a,Sanna18},
and Au/Ge(100)~\cite{Lichtenstein19},
as well as in ultrathin metallic silicide wires~\cite{Rugeramigabo10}.
The dispersions of plasmons is often investigated 
in relation to the dimensionality issue because
their long-wave-length dispersion within the Fermi liquid theory depends on the dimension,
\begin{equation}
\label{2dplasmon}
E(\vec{q}) \propto \sqrt{\vert \vec{q} \vert }
\end{equation}
in an isotropic 2D metal~\cite{Stern1967}
and  
\begin{equation}
\label{1dplasmon}
E(q) \propto \vert q \vert 
\end{equation}
in a 1D metal~\cite{Bruus04}.
However, the theoretical predictions for anisotropic 2D metals or coupled wires
are not so simple and clear-cut~\cite{Williams74,Gold92,Nagao06,Moudgil10,Schulz83,Kopietz95,Kopietz97}
and the experimental data rarely allow us to determine the behavior
in the long-wave-length limit $q=\vert \vec{q} \vert  \rightarrow 0$ with certainty. 

In this paper we discuss the dispersion of low-energy collective charge excitations (plasmons, holons)
in atomic wire systems
using well-established theoretical approaches and results: 
random phase approximation (RPA) for anisotropic Fermi liquids~\cite{,Bruus04,solyom3} and 
bosonization for coupled TLLs~\cite{Giamarchi03,Schulz83}
 as well as Bethe Ansatz~\cite{essler05} and density-matrix renormalization group~\cite{scho05,scho11,jeck08a}
methods for correlated ladder models of coupled chains.
In particular, we show that there is no clear-cut qualitative difference between the theoretical predictions
for strongly anisotropic 2D Fermi liquids and coupled 1D TLLs in the long wave length limit.
Moreover, in both approaches the behavior of $E(\vec{q})$ for $q \rightarrow 0$
reflects the screening of the Coulomb interaction between electrons rather than the dimensionality 
of the metallic state.
Additionally, our results suggest that low-energy charge excitations can exhibit a significant 2D dispersion,
even when the system is a strongly anisotropic (quasi-1D) conductor or exhibits a TLL power-law behavior in the density of states.
Thus it is possible that atomic wire systems such as Au/Ge(100)
possess a mixture of properties associated with 1D and 2D metals, as found
experimentally~\cite{Meyer11,Dudy17,Blumenstein11a,Meyer14,Nakatsuji11,Nakatsuji12,Park14,DeJong16,Lichtenstein19}.

\section{Strongly anisotropic Fermi liquids}

The theoretical properties of plasmons in low-dimensional metals are well understood
within a Fermi liquid approach.
 In particular, the dynamical responses of 2D metals~\cite{Stern1967}
 and  quasi-1D metals~\cite{Williams74,Gold92}  were
investigated several decades ago.
The dynamical response in quantum wires was compared with isotropic 2D  systems within the Fermi liquid theory~\cite{DasSarma96} and with the TLL theory~\cite{Wang01} using continuum models.
More recently, plasmon properties have been studied beyond RPA 
within the Fermi liquid theory~\cite{Nagao06,Moudgil10}.
Here we discuss some properties of plasmons in low-dimensions on a anisotropic lattice to facilitate the comparison
with the TLL theory for coupled chains and the numerical results for correlated ladder models in the next sections.

We consider a tight-binding system on a rectangular lattice with the lattice constant $a$
in the wire direction ($x$-direction) and a distance $b$ between wires ($y$-direction).
The hopping term between nearest-neighbor sites is denoted $t_{\parallel}$ in the wire direction and $t_{\perp}$ 
between wires. 
The system  can be seen as an anisotropic 2D lattice with $L_x \times L_y$ sites or as an array of $L_y$ chains
with $L_x$ sites.
In addition we take into account an  electron-electron interaction 
$V(\vec{r}) $ in the plane formed by the wires.
The Hamiltonian of the system is
\begin{eqnarray}
\label{hamiltonian}
H & = & - t_{\parallel} \sum_{x,y,\sigma} \left ( c ^{\dagger}_{x,y,\sigma} c^{\phantom{\dagger}}_{x+1,y,\sigma} + \text{h.c.} \right )
\\
&& -  t_{\perp} \sum_{x,y,\sigma} \left (  c ^{\dagger}_{x,y,\sigma} c^{\phantom{\dagger}}_{x,y+1,\sigma} + \text{h.c.}
\right )  \nonumber \\
&& + \sum_{\vec{r_1},\vec{r_2},\sigma_1,\sigma_2}  V(\vec{r}_1-\vec{r}_2) \  n_{\vec{r_1},\sigma_1} n_{\vec{r_2},\sigma_2} . \nonumber
\end{eqnarray}
The operator $c ^{\dagger}_{x,y,\sigma}$ creates an electron with spin $\sigma$ in the site with position
$\vec{r}=(xa,yb)$. $n_{\vec{r},\sigma}=c ^{\dagger}_{x,y,\sigma} c^{\phantom{\dagger}}_{x,y,\sigma}$ is the local particle number operator.
The first two sums run over all indices $x=1,\dots,L_x$, $y=1,\dots,L_y$, and $\sigma=\uparrow, \downarrow$,
while the third sum is over all pairs of sites. 

We determine the dispersion of plasmons using the RPA within the Fermi liquid theory. 
More precisely, 
we compute the first-order response of the electron gas to a dynamical
external electric field using a time-dependent Hartree-Fock approximation.
The dispersion of long-live collective charge excitations (plasmons)
is given by the vanishing of the real part of the Lindhard dielectric function~\cite{solyom3}.
We discuss only the results for long wave lengths ($q \rightarrow 0$)
 in the thermodynamic limit $L_x,L_y \rightarrow \infty$.
 
 We first consider  an isotropically screened Coulomb potential
\begin{equation}
\label{coulomb}
V(r) = \frac{e^2}{4\pi\epsilon} \frac{e^{-r/\xi}}{r}
\end{equation}
with screening length $\xi$, effective dielectric constant $\epsilon$, and electron charge $e$.
 For an isotropic lattice ($t_{\perp}=t_{\parallel}$, $a=b$, and $L_x=L_y$)  in the low-density regime
 we obtain the plasmon dispersion
\begin{equation}
\label{isotropicFermi}
E(\vec{q}) =  A \  q  \left ( \xi^{-2}+q^2 \right )^{-1/4}
\end{equation}
with the constant prefactor
\begin{equation}
A = \sqrt{\frac{e^2 t_{\parallel} a^2 n}{\epsilon} }
\end{equation}
where $n$ is the electron surface density. Assuming no screening ($\xi \rightarrow \infty$) and using the relation between the hopping term on a 2D lattice 
and the (renormalized) electron mass $m$ of the 2D Fermi gas in the continuum  [$t_{\parallel}a^2 = \hbar^2 / (2m)$],
we recover the known result for the plasmon dispersion in an isotropic 2D metallic system~\cite{Stern1967}
\begin{equation}
E(\vec{q}) = \hbar \sqrt{\frac{e^2 n}{2 \epsilon m} } \sqrt{q} .
\end{equation}
In a strongly anisotropic lattice, where the Fermi velocity in the wire direction  $v_{\text F} \propto t_{\parallel} a \gg t_{\perp} b$, we obtain the plasmon dispersion
\begin{equation}
\label{anisotropicFermi}
E(\vec{q}) =  B \sqrt{q_x^2 + R q_y^2}  \left ( \xi^{-2}+q^2 \right )^{-1/4}
\end{equation} 
with
\begin{equation}
B = \sqrt{\frac{e^2\hbar v_{\text F}}{\pi \epsilon b}}
\end{equation} 
and the dimensionless anisotropy parameter
\begin{equation}
R =  2 \frac{t_{\perp}^2 b^2}{\hbar^2  v_{\text F}^2} \ll 1.
\end{equation} 
In the low-density strongly-anisotropic  limit $\hbar v_{\text F} \approx \pi n t_{\parallel} a^2 b$ and thus 
$B=A$. Although, we have derived Eq.~(\ref{anisotropicFermi}) using the condition $R \ll 1$, 
we note that
it agrees with the isotropic case~(\ref{isotropicFermi}) if we set $R=1$.

In the absence of chain hybridization ($t_{\perp} = 0 \Rightarrow R=0$), the charge carriers can move
only in the wire direction and thus the system is a purely unidirectional conductor.
The plasmon dispersion is then $E(q_x)  \propto \vert q_x \vert$ for $q_x \rightarrow 0$ at any finite screening length $\xi$
and fixed $q_y$. This behavior appears to agree at least qualitatively with the result~(\ref{1dplasmon}) for a 1D metal
but the prefactor in Eq.~(\ref{anisotropicFermi}) 
is different from the result for a single wire~\cite{Bruus04} and depends on the normal component of the wave vector $q_y$ 
because the wires are still coupled by the 2D Coulomb potential
in our model. This interpretation is incorrect, however. 
Experimentally, the dispersion is measured as a function of the wave length $\lambda =2\pi/q$ either angle-resolved or averaged over all directions in the surface.
The theoretical dispersion must then be written 
\begin{equation}
\label{angle-resolved}
E(\vec{q}) = C(q)  \vert q_x \vert  = C(q) q \cos(\theta)
\end{equation}
with
\begin{equation}
C(q) = B  \left ( \xi^{-2}+q^2 \right )^{-1/4}
\end{equation}
where $\theta \in [0,\frac{\pi}{2}]$ is the angle between the wave vector $\vec{q}$ and the wire direction $x$.
Thus we recover  the typical angle-dependent plasmon frequency of a quasi-1D
metal~\cite{Williams74,Gold92}.
In contrast to isotropic materials,  there is a continuum of plasmon excitations
between the vanishing energy $E(\vec{q}) = 0$ for $\theta = \pi/2$ and
the maximal energy $E(\vec{q})=C(q) q$ for  $\theta=0$. 
For a fixed direction $\theta \neq \pi/2$
we see that the dispersion~(\ref{angle-resolved}) scales with the norm $q$ of the wave vector 
as predicted for 1D metals, Eq. (\ref{1dplasmon}), when the Coulomb interaction is screened (i.e., $\xi$ is finite)
but as predicted for isotropic 2D metals, Eq. (\ref{2dplasmon}), in the absence of screening ($\xi \rightarrow \infty$),
although the system conducts only in the wire direction in both cases.

This behavior is not an artifact of the vanishing interchain hopping. If $t_{\perp} \neq 0 \ (\Rightarrow R \neq 0)$,
the system is an anisotropic 2D conductor.  The dispersion of the Fermi wave vector  at the Fermi energy
 has a width $\Delta q_y = 4 t_{\perp}/(\hbar v_{\text F})$ in the  strongly anisotropic limit. 
We must similarly write the plasmon dispersion~(\ref{anisotropicFermi}) as a function of the angle $\theta$
\begin{equation}
E(\vec{q}) =  B \sqrt{\cos^2(\theta) + R \sin^2(\theta)}  \ q \left ( \xi^{-2}+q^2 \right )^{-1/4} .
\end{equation} 
Again we find that the dispersion scales as~(\ref{1dplasmon})  when the Coulomb interaction is screened
and as~(\ref{2dplasmon})  in the absence of screening ($\xi \rightarrow \infty$), although the system is
a 2D conductor with an anisotropic metallic dispersion at the Fermi energy in both cases. Therefore, the dispersions of plasmons in anisotropic metals do not characterize
their dimensionality but depends on the screening of the interaction between electrons.

This result can be generalized to other potential shapes. 
For instance, the screened Coulomb potential~(\ref{coulomb}) does not result in the non-monotonic plasmon dispersions observed 
experimentally  in Au/Ge(100)~\cite{Lichtenstein19}.
To reproduce the experimental curvature, the 2D Fourier transform of the 
interaction potential
\begin{equation}
\tilde{V}(\vec{q}) = \int V\left (\vec{r} \right ) e^{-i\vec{q}\cdot \vec{r}} d^2r 
\end{equation}
must decrease rapidly with increasing $q$ beyond some cutoff wave number $q_\text{c}$.
In Ref.~\cite{Lichtenstein19} a phenomenological isotropic gaussian potential was considered
\begin{equation}
\label{gaussian}
V(r) = V_0 e^{-r^2/(2\xi^2)}
\end{equation}
with $q_\text{c}= \sqrt{2}/\xi$. This results in the plasmon dispersion
\begin{equation}
E(\vec{q}) =  B' \sqrt{\cos^2(\theta) + R \sin^2(\theta)}  \ q \ e^{-q^2\xi^2/2}
\end{equation} 
with 
\begin{equation}
B' = \sqrt{\frac{4\hbar V_0 \xi^2 v_{\text F}}{b}}.
\end{equation} 
The comparison with the experimental data is discussed in the next section.
Here we just want to point out that for this gaussian  potential, as for all interaction potentials
$\tilde{V}(\vec{q})$ that remain finite  for $\vec{q} \rightarrow 0$, the plasmon dispersion
scales as in 1D metals, Eq.~(\ref{1dplasmon}), in the long-wave-length limit, irrespective of whether
the system conducts in one direction ($t_{\perp}=0$) or in two directions ($t_{\perp}\neq 0$).
Actually, the fact that the energy of collective density oscillations is proportional to $q$ for short-range interactions
in any dimension
is well known  since Landau's Fermi liquid theory of the
zero sound in $^3$He~\cite{solyom3}.

\section{Coupled Tomonaga-Luttinger liquids}

The theory of low-energy excitations in strongly correlated  systems of coupled metallic 
chains is not so well developed as for purely 1D metals (TLLs) and Fermi liquids.
Is it established, however, that two-body interactions between electrons
 lead to a Fermi liquid or an insulating state for any finite interchain hopping
 but the system may remain a TLL for vanishing interchain hopping~\cite{Schulz83,Kopietz95,Kopietz97,Giamarchi03}.
 Here we use and compare two approaches for correlated wire systems without hybridization ($t_{\perp}=0$):
 bosonization for broad systems with linearized bare dispersions
 and DMRG for two-leg ladder systems.

\subsection{Bosonization}

We first consider the generalized Tomonaga-Luttinger model introduced
by Schulz for a three-dimensional array of 1D conductors with an unscreened Coulomb potential~\cite{Schulz83}.
Only the forward scattering for small momentum transfer is considered explicitly and thus the model of coupled chains
can be solved using bosonization.
We have adapted this study to the case of a 2D array of wires with a general
electron-electron interaction $V(\vec{r})$. 
(Note that we use  the notation of Ref.~\cite{Giamarchi03}).
The system is a 2D array of 1D conductors with linear bare dispersions.
Electrons can move freely along a wire ($x$-direction) with a Fermi velocity $v_{\text F}$ but
perpendicular motion ($y$-direction) is forbidden.
The interaction acts both between electrons in the same chain and in different chains.
This generalized Tomonaga-Luttinger model  corresponds to the weak-coupling limit of the Hamiltonian~(\ref{hamiltonian})
with $t_{\perp}=0$. In particular, the system conducts charge in the wire direction only.

Following Ref.~\cite{Schulz83} we find that 
the dispersion of holons (collective charge excitations or equivalently plasmon) is
\begin{equation}
E(\vec{q}) = \hbar u(\vec{q}) \vert q_x \vert 
\end{equation}
where
the velocity $u(\vec{q})$ of elementary charge excitations in the wire direction  is given by
\begin{equation}
\label{holon}
\frac{u(\vec{q})^2}{v_{\text F}^2} = K^{-2}(\vec{q}) = 1 + \frac{2}{\pi \hbar
v_{\text F} b} \tilde{V}(\vec{q}) 
\end{equation}
with the dimensionless Luttinger liquid parameter $K(\vec{q})$.
Note that in this approach  $\tilde{V}(\vec{q})$ is assumed to be the Fourier transform of the long-range part of the interaction between electrons
while $v_{\text F}$ is the charge velocity of the interacting 1D conductors without this  long-range part of the interaction.
Thus $v_{\text F}$ may already be renormalized by the short-range interactions
within a single wire~\cite{Schulz83}.
For an isotropic 2D interaction $[V(\vec{r}) = V(r)]$ the dispersion can be written
\begin{equation}
\label{TL-angle-resolved}
E(\vec{q}) = \hbar u(q) q_x  =\hbar u(q) q \cos(\theta).
\end{equation}
Thus we recover the angle dependence~(\ref{angle-resolved}) found in the RPA calculation 
but the function $C(q)$ and $\hbar u(q)$ are different.

To illustrate this general result we again consider  the isotropically screened Coulomb potential~(\ref{coulomb}).
The resulting plasmon dispersion is 
\begin{equation}
\label{coulomb2d}
E(\vec{q}) = \hbar v_{\text F} \cos(\theta) q
\left [ 1 + \frac{D}{\sqrt{1+q^2 \xi^2}}  \right ]^{\frac{1}{2}}
\end{equation}
with the dimensionless constant
\begin{equation}
\label{D1}
D = \frac{e^2 \xi}{\pi \epsilon b \hbar v_{\text F}}.
\end{equation}
This result reveals the essential qualitative difference between 
the RPA~(\ref{anisotropicFermi}) and TLL predictions for the plasmon and holon dispersions, respectively.
RPA predicts erroneously that there are no long-live collective charge excitations in a 1D conductor 
in the absence of the electron-electron interaction [i.e., there is no solution $E(\vec{q}) \neq 0$
for $e^2/\epsilon = 0$]. The TLL theory shows correctly that collective excitations
exist in a 1D conductor even in the absence of interactions.
The point at issue in this work is the scaling for long wave lengths, however.
The dispersion~(\ref{coulomb2d}) scales for $q \rightarrow 0$ as in a 1D metal, Eq.~(\ref{1dplasmon}), 
for a screened Coulomb interaction (finite $\xi$) but as in a 2D metal,  Eq.~(\ref{2dplasmon}),
without screening ($\xi \rightarrow \infty$). 
Therefore, there is no qualitative difference between RPA and TLL  theory regarding the dispersion
for small $q$.
Using the relation for an isotropic 2D electron gas
\begin{equation}
v_{\text F}  = \frac{\pi \hbar b n}{2m}
\end{equation}
we even find that the RPA plasmon dispersion~(\ref{angle-resolved}) and the
TLL holon dispersion~(\ref{coulomb2d}) are exactly equal
\begin{equation}
E(\vec{q}) = \hbar \sqrt{\frac{e^2 n}{2 m \epsilon_0}} \sqrt{q} \cos(\theta) 
\end{equation}
in the absence of screening ($\xi \rightarrow \infty$).

\begin{figure}
\centering
\includegraphics[width=0.99\columnwidth]{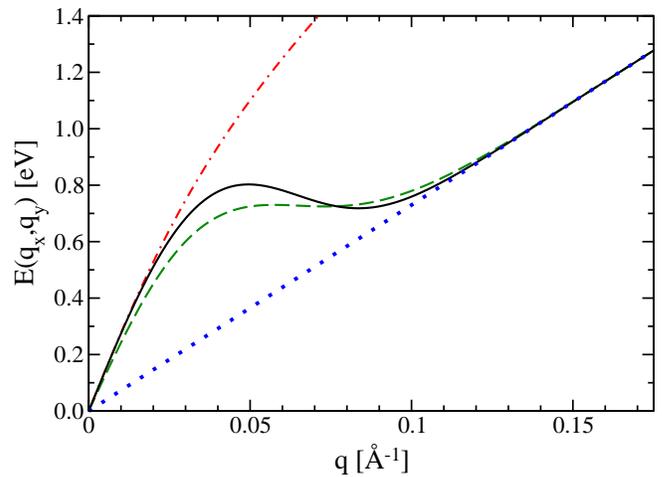}
\caption{\label{fig1} Dispersions $E(\vec{q})$ of the upper edge of the continuum ($\theta=0$) of collective charge excitations 
(holons) in coupled TLLs as a function of $q=\vert\vec{q}\vert$.  The red dash-dotted curve shows the 
dispersion~(\ref{coulomb2d}) for a screened Coulomb potential.
The solid black curve indicates the dispersion~(\ref{gaussian2d}) for a phenomenological gaussian potential.
The blue dotted curve corresponds to uncoupled TLLs.
The dashed green curve represents a fit to the experimental data for plasmons in Au/Ge(100) presented in Ref.~\cite{Lichtenstein19}.}
\end{figure}

The upper edge of the dispersion~(\ref{coulomb2d}) is plotted  in Fig.~\ref{fig1} for parameters corresponding to
gold wires on a Ge(100) surface.
The interpretation of STS data with the TLL theory~\cite{Blumenstein11a}
yields $K(0)= 1/\sqrt{1 + D}\approx 0.26$ and thus $D\approx 13.8$.
From electron energy loss spectroscopy~\cite{Lichtenstein19} we get $v_{\text F} \approx 1.1 \times 10^6$ ms$^{-1}$
or $\hbar v_{\text F} \approx 7.3$~eV\AA.  If we choose the dielectric constant of vacuum  $\epsilon=\epsilon_0$,
we then get a large screening length $\xi \approx 27.6$ \AA $\ \gg a,b$ from Eq.~(\ref{D1}).
The continuum of holon excitations~(\ref{coulomb2d})  extends from the horizontal axis up to this curve as $\theta$ varies.
The  experimental data are represented  in Fig.~\ref{fig1} by a fitted theoretical curve (see Ref.~\cite{Lichtenstein19}).
This curve is indeed within the theoretical boundaries of the continuum but clearly
the screened Coulomb potential~(\ref{coulomb}) does not result in the non-monotonic plasmon dispersions observed 
experimentally  in Au/Ge(100).

To reproduce the experimental curvature, the potential $\tilde{V}(\vec{q})$ must
decreases rapidly with increasing $q$ beyond some cutoff wave number 
as with the phenomenological gaussian potential~(\ref{gaussian}).
The plasmon dispersion is then
\begin{equation}
\label{gaussian2d}
E(\vec{q}) = \hbar v_{\text F}  \cos(\theta)  q
\left [ 1 + D' \exp\left(-\frac{q^2 \xi^2}{2}\right)  \right ]^{\frac{1}{2}}
\end{equation}
with the dimensionless constant
\begin{equation}
\label{D2}
D' = \frac{4 V_0 \xi^2}{b \hbar v_{\text F}}.
\end{equation}
We again use the experimental values  $v_{\text F}$ and $K(0)=1/\sqrt{1+D'}$ mentioned above
for the screened Coulomb potential. The remaining free parameter is set
to $\xi = 3.2$ nm to reproduce the experimentally observed curvature.
This screening length is twice as large as the distance 
$b=1.6$ nm between gold wires on the germanium surface
according to Ref.~\cite{Blumenstein11a},
This also determines the potential strength $V_0 \approx 0.39$ eV.
This value is consistent with a strongly screened Coulomb interaction
at length scales larger than the interchain distance $b$
because the Coulomb energy between two electrons at distance
$b$ is $e^2/(4\pi\epsilon_0 b) \approx 0.9$ eV.
The upper edge of the holon  dispersion~(\ref{gaussian2d})
is shown in Fig.~\ref{fig1}  and compared to the curve deduced
from the experimental data for plasmons in Au/Ge(100)~ \cite{Lichtenstein19}.
The agreement between the theoretical and experimental dispersions
is satisfactory. As noted  in Ref.~\cite{Lichtenstein19}, however, the value  of the velocity $v_{\text F}$ is incompatible with the
value obtained from photoemission experiments~\cite{Meyer11,DeJong16}.
Additionally, the short fitted screening length $\xi$ is not fully consistent with the assumptions made 
to compute the dispersion~(\ref{holon}) within the TLL  theory.
Clearly, the holon dispersion~(\ref{gaussian2d})  is not monotonic with increasing $q$ as illustrated in Fig.~\ref{fig1}.
 Non-monotonic  dispersions for collective charge excitations seem to be a generic phenomenon in 
 1D electron systems with long-range interactions~\cite{Chou2020}.

Both examples show that the electron-electron interaction between TLL wires induces a significant 
dispersion of the holon energies $E(\vec{q})$ as a function of the perpendicular component $q_y$ of the wave vector.
As all dynamical response functions involving charge excitations are derived from 
these elementary excitations, their dispersion can exhibit
a 2D character, although the system conducts in the wire direction only. 
This result agrees qualitatively with the observations made for plasmons using RPA in the previous section.
Therefore, both the TLL and Fermi liquid approaches suggest that  the momentum-dependence of response functions 
in strongly anisotropic 2D conductors ($v_{\text F} \gg b t_{\perp} \neq 0$)   
could be determined by the strength of the inter-wire electron-electron interaction rather than  the amplitude of the inter-chain hopping.
As a consequence,  a quantity like the single-particle Green's  function, which corresponds to the spectrum measured in photoemission experiments,
could  exhibit a significant 2D dispersion at the Fermi energy even if the system remains a strongly anisotropic (i.e., quasi-1D) conductor.

\subsection{DMRG for ladder systems}

To obtain additional information we have investigated the correlated lattice model~(\ref{hamiltonian}) numerically on a two-leg lattice
using the density-matrix renormalization group (DMRG) method~\cite{scho05,scho11,jeck08a}. The dynamical charge structure factor
is defined by 
\begin{equation}
\label{strucFac}
S(\vec{q},\omega) = \frac{1}{\pi} \text{Im}  \left \langle n_{-\vec{q},\sigma} \frac{1}{H-\hbar \omega-E_0-i\eta} n_{\vec{q},\sigma} \right \rangle
\end{equation}
where the expectation value is calculated for the many-body ground state of $H$, $E_0$ is its energy, $n_{\vec{q},\sigma}$ is
the Fourier transform of the local
particle number operator $n_{\vec{r},\sigma}$, and $\eta$ is a small positive number that broadens the spectrum.
In a TLL the function  $S(\vec{q},\omega)$ exhibits dispersive features $\hbar \omega(\vec{q})$ that are related to 
the holon excitation branches or a combination
 thereof~\cite{bent07}. Thus one can determine the holon dispersions from the dynamical charge structure factor.

For narrow quasi-1D correlated systems
$S(\vec{q},\omega)$ can be computed with the dynamical DMRG method~\cite{jeck02a,bent07}.
The computational cost is very high, however, and increases exponentially with the system width $L_y$.
Therefore, we restrict our DMRG study to a ladder system with $L_y=2$ and spinless fermions [i.e., all electrons have the same spin polarization
and thus we can drop the index $\sigma$ in the definitions of the Hamiltonian~(\ref{hamiltonian}) and the structure factor~(\ref{strucFac})].
Additionally, we will take into account only the nearest-neighbor interactions $V_x=V(\vec{r}=(a,0))$ in the wire direction and $V_y=V(\vec{r}=(0,b))$ 
between wires
as well as the diagonal next-nearest-neighbor interaction $V_{xy}=V(\vec{r}=(a,b))$.
As mentioned in the previous section, the hopping term $t_{\perp}$ leads rapidly to insulating phases (e.g., charge-density-wave  ground states),
thus we consider only the case $t_{\perp}=0$.

This simplified model can be mapped exactly onto a 1D extended $U-V$ Hubbard model for electrons when $V_x = V_{xy}$. 
The local interaction (Hubbard term) is then $U=V_y$ while the nearest-neighbor interaction is $V=V_x = V_{xy}$.
The ground-state phase diagram
and the Luttinger parameters of this model at quarter filling (i.e., with $N=L_x/2$ fermions) are well known~\cite{Mila1993,Penc1994,Ejima2005,Shirakawa2009} 
and thus we can easily find model parameters corresponding  to a TLL phase.
Moreover, the Hubbard model ($V=0$) is exactly solvable using the Bethe Ansatz~\cite{essler05} and thus we can compute the holon dispersions
in the simplified Hamiltonian~(\ref{hamiltonian}) exactly in that case.

\begin{figure}
\centering
\includegraphics[width=0.99\columnwidth]{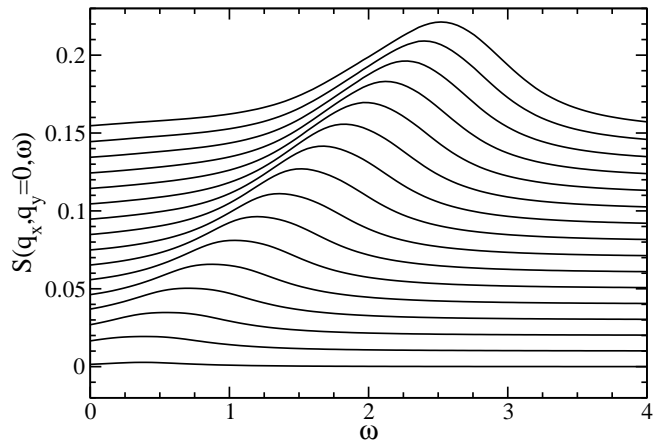}
\caption{\label{fig2} 
Dynamical charge structure factor $S(\vec{q},\omega)$~(\ref{strucFac}) calculated with DMRG for a two-leg ladder with 
$V_y=8t_{\parallel}$, $V_x=V_{xy}=0$ and $q_y=0$ as function of the excitation energy $\hbar \omega$
for several values of $q_x$ from $\pi/33$ (bottom) to $16\pi/33$ (top).
The system length is $L_x=32$ and the broadening is $\eta=0.4 t_{\parallel}$.
The units are $t_{\parallel}=1$ and $\hbar=1$.
}
\end{figure}

We carry out DMRG computations using up to 800 density-matrix eigenstates, resulting
in discarded weights smaller than $10^{-6}$.
The system sizes range from $L_x=32$ to $L_x=128$
with a broadening $\eta/t_{\parallel} = 0.1$ to $0.4$.
We use open boundary conditions and pseudo wave numbers $q_x=z \pi/(L_x+1)$ with $z=1,\dots,L_x$  and $q_y=0,\pi$ because momentum-resolved dynamical 
DMRG simulations are simpler with this choice~\cite{bent07}.
We have found as expected that most of the spectral weight of $S(\vec{q},\omega)$ is located close to $\vert q_x\vert =\pi/a$ and $\pi/(2a)$.  
This is the signature of the 
1D quasi-long-range charge-density-wave order with wave number $2k_{\text F}$ and $4k_{\text F}$. Nevertheless, we are able
to determine the spectrum and the holon dispersions for smaller $\vert q_x \vert$ accurately
because $S(\vec{q},\omega)$ is calculated separately for each wave vector $\vec{q}$ with the dynamical DMRG method.
Figure~\ref{fig2} shows an example of the calculated spectrum $S(\vec{q},\omega)$ for $0 \alt q_x \alt 2k_{\text F} = \pi/(2a) $. The position of the maxima
as a function of the excitation energy $\omega$ for each wave vector $\vec{q}$ yields the holon dispersion
$E(\vec{q}) = \hbar \omega(\vec{q})$.
The accuracy of the resulting data is limited by the spectrum broadening $\eta$ for the energy $\hbar \omega$ and by the discretization
$\pi/(L_x+1)$ for the wave vector.

\begin{figure}
\centering
\includegraphics[width=0.99\columnwidth]{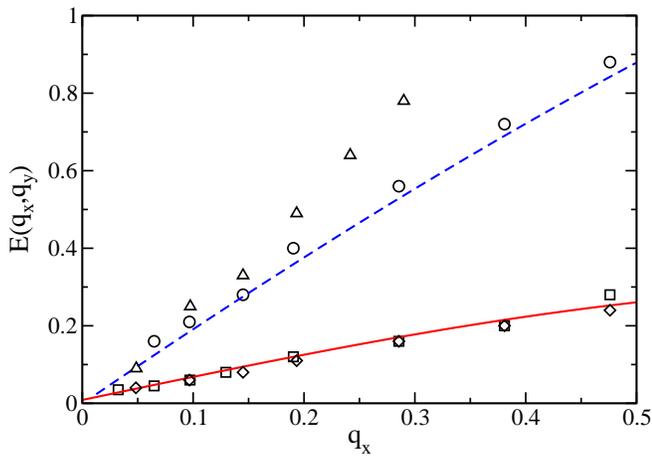}
\caption{\label{fig3}  Holon dispersions $E(\vec{q})$  in two-leg ladders as a function of $q_x$ for fixed $q_y$. 
Symbols show values determined from the charge structure factor calculated with DMRG
for $V_y=8t_{\parallel}$,  $V_x=V_{xy}=0$, $q_y=0$ (circles) and $q_y=\pi/a$ (squares) as well as
for $V_x=V_y=V_{xy}=4t_{\parallel}$,  $q_y=0$ (triangles) and $q_y=\pi/a$ (diamonds). 
The Bethe Ansatz solutions for $V_y=8t_{\parallel}$,  $V_x=V_{xy}=0$ are represented by a solid red line ($q_y=\pi/a$) and a blue dashed line
($q_y=0$), respectively.
The units are $t_{\parallel}=1$ and $a=1$.
}
\end{figure}

Figure~\ref{fig3} shows two examples of the holon dispersions obtained from the structure factor.
First, we see that our numerical results for $V_y=8t_{\parallel}$ and $V_x=V_{xy}=0$ agree very well
with the exact dispersions calculated from  the Bethe Ansatz solution. The second example 
corresponds to an isotropic interaction $V_x=V_y=V_{xy}=4t_{\parallel}$.  The similitude of the dispersion
for $q_y=\pi/a$ with the Bethe Ansatz solution is a coincidence.
The exact Bethe Ansatz dispersions  are linear for $q_x \rightarrow 0$, 
\begin{equation}
E(\vec{q}) =  F(q_y) \vert q_x \vert.
\end{equation}
Although one clearly observes a curvature at finite $q_x$, our DMRG data are compatible with this linear dispersion in
the limit $q_x \rightarrow 0$ for other interaction  parameters $V_x,V_y,V_{xy}$
leading to a TLL ground state.
Moreover, we observe in Fig.~\ref{fig3} that the dispersions, particularly the slopes $F(q_y)$, strongly depend on the normal component $q_y$
of the wave vector.
Therefore, our numerical  data agree with the  generic holon dispersion~(\ref{TL-angle-resolved}) predicted by the TLL theory
for system of 1D conductors. In particular, they confirm that the holon dispersion can be significant in the direction perpendicular
to the wires, even when the system is a unidirectional conductor.

On the other hand, it is well known that the local density of states (LDOS) of 1D TLLs exhibits a power-law behavior at the Fermi energy~\cite{Giamarchi03}.
This behavior has been observed explicitly in correlated 1D lattice models similar to the ones studied here using
numerical methods~\cite{Meden2000,Andergassen2006,jeck13}
but this requires much longer system lengths than those used in the present work. Nevertheless, it is certain that the power-law scaling of the LDOS occurs
in the two-leg ladder TLL studied here.
Therefore, our investigation of the Hamiltonian~(\ref{hamiltonian}) for coupled TLLs suggests that
one can observe both a significant 2D dispersion of elementary charge excitations and a TLL power-law behavior of the LDOS at the Fermi energy.
This could explain some of the apparently conflicting experimental results for gold chains on germanium
surfaces~\cite{Meyer11,Dudy17,Blumenstein11a,Meyer14,Nakatsuji11,Nakatsuji12,Park14,DeJong16,Lichtenstein19}.
Naturally, investigations of the single-particle spectral functions and the LDOS in broader systems of coupled chains ($L_y > 2$) are necessary to confirm these findings,
but they are too computationally expensive currently.

\section{Conclusions}

We have investigated the low-energy collective charge excitations (plasmons, holons)
in strongly anisotropic 2D lattices or weakly coupled wires with a view 
to understanding metallic states in 2D arrays of atomic wires deposited on semiconducting substrates.
Various aspects  have been neglected and most results have been obtained using approximate methods.
For instance, it is known that the substrate modifies the effective interaction between conduction electrons in the wires and
thus influences the properties of TLL ~\cite{Abdelwahab18a}.
Nevertheless, three main findings arise from the present study.
First, the Fermi liquid  and TLL  theories predict the same qualitative behavior
for the dispersion $E(\vec{q})$ of these excitations for long wave lengths.
Second, their scaling for $q \rightarrow 0$ depends on the choice of the effective electron-electron interaction
but does not characterize the dimensionality of the metallic state.
Third, the same system can exhibit a 2D dispersion of low-energy excitations due to the coupling between wires
but remain a strongly anisotropic conductor or retain typical features of a TLL such as the power-law behaviour of the LDOS at the Fermi energy.
Therefore, we are not able to propose a practical criterion to distinguish
between a strongly anisotropic 2D Fermi liquid and  a system of weakly-coupled TLL wires.
Actually, it is probable  that metallic states in real atomic wire materials  possess properties of both 2D and 1D metals
that can be revealed in different experiments, as suggested by the diverse features found for Au/Ge(100).

\begin{acknowledgments}
This work was done
as part of the Research Unit \textit{Metallic nanowires on the atomic scale: Electronic
and vibrational coupling in real world systems} (FOR1700) 
of the German Research Foundation (DFG) and was supported by
grant No.~JE~261/1-2.
\end{acknowledgments}

\bibliographystyle{biblev1}
\bibliography{mybibliography}{}

\begin{thebibliography}{10}
\expandafter\ifx\csname url\endcsname\relax
  \def\url#1{{\tt #1}}\fi
\expandafter\ifx\csname urlprefix\endcsname\relax\def\urlprefix{URL }\fi
\expandafter\ifx\csname bibinfo\endcsname\relax\def\bibinfo#1#2{#2}\fi
\expandafter\ifx\csname eprint\endcsname\relax\def\eprint#1{\url{#1}}\fi

\bibitem{Springborg07}
\bibinfo{author}{M.~Springborg} and \bibinfo{author}{Y.~Dong}, {\em
  \bibinfo{title}{Metallic Chains/Chains of Metals}\/}
  (\bibinfo{publisher}{Elsevier}, \bibinfo{address}{Amsterdam},
  \bibinfo{year}{2007}).

\bibitem{Oncel08}
\bibinfo{author}{N.~Oncel}, \bibinfo{title}{Atomic chains on surfaces},
  \bibinfo{journal}{\href{http://dx.doi.org/10.1088/0953-8984/20/39/393001}{J.
  Phys.: Condens. Matter}}
  \href{http://dx.doi.org/10.1088/0953-8984/20/39/393001}{{\bf
  \bibinfo{volume}{20}}, \bibinfo{pages}{393001}}
  (\href{http://dx.doi.org/10.1088/0953-8984/20/39/393001}{\bibinfo{year}{2008}}).

\bibitem{Dudy17}
\bibinfo{author}{L.~Dudy}, \bibinfo{author}{J.~Aulbach},
  \bibinfo{author}{T.~Wagner}, \bibinfo{author}{J.~Schäfer}, and
  \bibinfo{author}{R.~Claessen}, \bibinfo{title}{One-dimensional quantum
  matter: gold-induced nanowires on semiconductor surfaces},
  \bibinfo{journal}{\href{http://dx.doi.org/10.1088/1361-648x/aa852a}{J. Phys.:
  Condens. Matter}} \href{http://dx.doi.org/10.1088/1361-648x/aa852a}{{\bf
  \bibinfo{volume}{29}}, \bibinfo{pages}{433001}}
  (\href{http://dx.doi.org/10.1088/1361-648x/aa852a}{\bibinfo{year}{2017}}).

\bibitem{Giamarchi03}
\bibinfo{author}{T.~Giamarchi}, {\em \bibinfo{title}{Quantum Physics in One
  Dimension}\/}, International Series of Monographs on Physics
  (\bibinfo{publisher}{Clarendon Press}, \bibinfo{address}{Oxford},
  \bibinfo{year}{2003}).

\bibitem{Bruus04}
\bibinfo{author}{H.~Bruus} and \bibinfo{author}{K.~Flensberg}, {\em
  \bibinfo{title}{Many-Body Quantum Theory in Condensed Matter Physics}\/}
  (\bibinfo{publisher}{Oxford University Press}, \bibinfo{address}{Oxford},
  \bibinfo{year}{2004}).

\bibitem{solyom3}
\bibinfo{author}{J.~{S\'{o}lyom}}, {\em \bibinfo{title}{Fundamentals of the
  Physics of Solids, Volume 3 - Normal, Broken-Symmetry, and Correlated
  Systems}\/} (\bibinfo{publisher}{Springer}, \bibinfo{address}{Berlin},
  \bibinfo{year}{2010}).

\bibitem{Meyer11}
\bibinfo{author}{S.~Meyer}, \bibinfo{author}{J.~Sch{\"{a}}fer},
  \bibinfo{author}{C.~Blumenstein}, \bibinfo{author}{P.~H{\"{o}}pfner},
  \bibinfo{author}{A.~Bostwick}, \bibinfo{author}{J.~L. McChesney},
  \bibinfo{author}{E.~Rotenberg}, and \bibinfo{author}{R.~Claessen},
  \bibinfo{title}{{Strictly one-dimensional electron system in Au chains on
  Ge(001) revealed by photoelectron k-space mapping}},
  \bibinfo{journal}{\href{http://dx.doi.org/10.1103/PhysRevB.83.121411}{Phys.
  Rev. B}} \href{http://dx.doi.org/10.1103/PhysRevB.83.121411}{{\bf
  \bibinfo{volume}{83}}, \bibinfo{pages}{121411}}
  (\href{http://dx.doi.org/10.1103/PhysRevB.83.121411}{\bibinfo{year}{2011}}).

\bibitem{Blumenstein11a}
\bibinfo{author}{C.~Blumenstein}, \bibinfo{author}{J.~Sch{\"a}fer},
  \bibinfo{author}{S.~Mietke}, \bibinfo{author}{S.~Meyer},
  \bibinfo{author}{A.~Dollinger}, \bibinfo{author}{M.~Lochner},
  \bibinfo{author}{X.~Y. Cui}, \bibinfo{author}{L.~Patthey},
  \bibinfo{author}{R.~Matzdorf}, and \bibinfo{author}{R.~Claessen},
  \bibinfo{title}{Atomically controlled quantum chains hosting a
  {T}omonaga-{L}uttinger liquid},
  \bibinfo{journal}{\href{http://dx.doi.org/10.1038/nphys2051}{Nat Phys}}
  \href{http://dx.doi.org/10.1038/nphys2051}{{\bf \bibinfo{volume}{7}},
  \bibinfo{pages}{776}}
  (\href{http://dx.doi.org/10.1038/nphys2051}{\bibinfo{year}{2011}}).

\bibitem{Meyer14}
\bibinfo{author}{S.~Meyer}, \bibinfo{author}{L.~Dudy},
  \bibinfo{author}{J.~Sch{\"a}fer}, \bibinfo{author}{C.~Blumenstein},
  \bibinfo{author}{P.~H{\"o}pfner}, \bibinfo{author}{T.~E. Umbach},
  \bibinfo{author}{A.~Dollinger}, \bibinfo{author}{X.~Y. Cui},
  \bibinfo{author}{L.~Patthey}, and \bibinfo{author}{R.~Claessen},
  \bibinfo{title}{Valence band and core-level photoemission of {Au}/{Ge}(001):
  {B}and mapping and bonding sites},
  \bibinfo{journal}{\href{http://dx.doi.org/10.1103/PhysRevB.90.125409}{Phys.
  Rev. B}} \href{http://dx.doi.org/10.1103/PhysRevB.90.125409}{{\bf
  \bibinfo{volume}{90}}, \bibinfo{pages}{125409}}
  (\href{http://dx.doi.org/10.1103/PhysRevB.90.125409}{\bibinfo{year}{2014}}).

\bibitem{Nakatsuji11}
\bibinfo{author}{K.~Nakatsuji}, \bibinfo{author}{Y.~Motomura},
  \bibinfo{author}{R.~Niikura}, and \bibinfo{author}{F.~Komori},
  \bibinfo{title}{Shape of metallic band at single-domain {Au}-adsorbed
  {Ge}(001) surface studied by angle-resolved photoemission spectroscopy},
  \bibinfo{journal}{\href{http://dx.doi.org/10.1103/PhysRevB.84.115411}{Phys.
  Rev. B}} \href{http://dx.doi.org/10.1103/PhysRevB.84.115411}{{\bf
  \bibinfo{volume}{84}}, \bibinfo{pages}{115411}}
  (\href{http://dx.doi.org/10.1103/PhysRevB.84.115411}{\bibinfo{year}{2011}}).

\bibitem{Nakatsuji12}
\bibinfo{author}{K.~Nakatsuji} and \bibinfo{author}{F.~Komori},
  \bibinfo{title}{Debate over dispersion direction in a
  {T}omonaga-{L}uttinger-liquid system},
  \bibinfo{journal}{\href{http://dx.doi.org/10.1038/nphys2240}{Nat. Phys.}}
  \href{http://dx.doi.org/10.1038/nphys2240}{{\bf \bibinfo{volume}{8}},
  \bibinfo{pages}{174}}
  (\href{http://dx.doi.org/10.1038/nphys2240}{\bibinfo{year}{2012}}).

\bibitem{Park14}
\bibinfo{author}{J.~Park}, \bibinfo{author}{K.~Nakatsuji},
  \bibinfo{author}{T.-H. Kim}, \bibinfo{author}{S.~K. Song},
  \bibinfo{author}{F.~Komori}, and \bibinfo{author}{H.~W. Yeom},
  \bibinfo{title}{Absence of {L}uttinger liquid behavior in {Au}-{Ge} wires:
  {A} high-resolution scanning tunneling microscopy and spectroscopy study},
  \bibinfo{journal}{\href{http://dx.doi.org/10.1103/PhysRevB.90.165410}{Phys.
  Rev. B}} \href{http://dx.doi.org/10.1103/PhysRevB.90.165410}{{\bf
  \bibinfo{volume}{90}}, \bibinfo{pages}{165410}}
  (\href{http://dx.doi.org/10.1103/PhysRevB.90.165410}{\bibinfo{year}{2014}}).

\bibitem{DeJong16}
\bibinfo{author}{N.~de~Jong}, \bibinfo{author}{R.~Heimbuch},
  \bibinfo{author}{S.~Eli{\"{e}}ns}, \bibinfo{author}{S.~Smit},
  \bibinfo{author}{E.~Frantzeskakis}, \bibinfo{author}{J.-S. Caux},
  \bibinfo{author}{H.~J.~W. Zandvliet}, and \bibinfo{author}{M.~S. Golden},
  \bibinfo{title}{{Gold-induced nanowires on the Ge(100) surface yield a 2D and
  not a 1D electronic structure}},
  \bibinfo{journal}{\href{http://dx.doi.org/10.1103/PhysRevB.93.235444}{Phys.
  Rev. B}} \href{http://dx.doi.org/10.1103/PhysRevB.93.235444}{{\bf
  \bibinfo{volume}{93}}, \bibinfo{pages}{235444}}
  (\href{http://dx.doi.org/10.1103/PhysRevB.93.235444}{\bibinfo{year}{2016}}).

\bibitem{Ohtsubo15}
\bibinfo{author}{Y.~Ohtsubo}, \bibinfo{author}{J.-i. Kishi},
  \bibinfo{author}{K.~Hagiwara}, \bibinfo{author}{P.~Le~F{\`e}vre},
  \bibinfo{author}{F.~Bertran}, \bibinfo{author}{A.~Taleb-Ibrahimi},
  \bibinfo{author}{H.~Yamane}, \bibinfo{author}{S.-i. Ideta},
  \bibinfo{author}{M.~Matsunami}, \bibinfo{author}{K.~Tanaka}, and
  \bibinfo{author}{S.-i. Kimura}, \bibinfo{title}{Surface
  {T}omonaga-{L}uttinger-liquid state on {$\mathrm{Bi}/\mathrm{InSb}(001)$}},
  \bibinfo{journal}{\href{http://dx.doi.org/10.1103/PhysRevLett.115.256404}{Phys.
  Rev. Lett.}} \href{http://dx.doi.org/10.1103/PhysRevLett.115.256404}{{\bf
  \bibinfo{volume}{115}}, \bibinfo{pages}{256404}}
  (\href{http://dx.doi.org/10.1103/PhysRevLett.115.256404}{\bibinfo{year}{2015}}).

\bibitem{Yaji16}
\bibinfo{author}{K.~Yaji}, \bibinfo{author}{S.~Kim},
  \bibinfo{author}{I.~Mochizuki}, \bibinfo{author}{Y.~Takeichi},
  \bibinfo{author}{Y.~Ohtsubo}, \bibinfo{author}{P.~L. F{\`{e}}vre},
  \bibinfo{author}{F.~Bertran}, \bibinfo{author}{A.~Taleb-Ibrahimi},
  \bibinfo{author}{S.~Shin}, and \bibinfo{author}{F.~Komori},
  \bibinfo{title}{One-dimensional metallic surface states of pt-induced atomic
  nanowires on {Ge}(001)},
  \bibinfo{journal}{\href{http://dx.doi.org/10.1088/0953-8984/28/28/284001}{J.
  Phys.: Condens. Matter}}
  \href{http://dx.doi.org/10.1088/0953-8984/28/28/284001}{{\bf
  \bibinfo{volume}{28}}, \bibinfo{pages}{284001}}
  (\href{http://dx.doi.org/10.1088/0953-8984/28/28/284001}{\bibinfo{year}{2016}}).

\bibitem{Hwang07}
\bibinfo{author}{C.~G. Hwang}, \bibinfo{author}{N.~D. Kim},
  \bibinfo{author}{S.~Y. Shin}, and \bibinfo{author}{J.~W. Chung},
  \bibinfo{title}{Possible evidence of non-{F}ermi liquid behaviour from
  quasi-one-dimensional indium nanowires},
  \bibinfo{journal}{\href{http://dx.doi.org/10.1088/1367-2630/9/8/249}{New
  Journal of Physics}} \href{http://dx.doi.org/10.1088/1367-2630/9/8/249}{{\bf
  \bibinfo{volume}{9}}, \bibinfo{pages}{249}}
  (\href{http://dx.doi.org/10.1088/1367-2630/9/8/249}{\bibinfo{year}{2007}}).

\bibitem{Liu08}
\bibinfo{author}{C.~Liu}, \bibinfo{author}{T.~Inaoka},
  \bibinfo{author}{S.~Yaginuma}, \bibinfo{author}{T.~Nakayama},
  \bibinfo{author}{M.~Aono}, and \bibinfo{author}{T.~Nagao},
  \bibinfo{title}{{Disappearance of the quasi-one-dimensional plasmon at the
  metal-insulator phase transition of indium atomic wires}},
  \bibinfo{journal}{\href{http://dx.doi.org/10.1103/PhysRevB.77.205415}{Phys.
  Rev. B}} \href{http://dx.doi.org/10.1103/PhysRevB.77.205415}{{\bf
  \bibinfo{volume}{77}}, \bibinfo{pages}{205415}}
  (\href{http://dx.doi.org/10.1103/PhysRevB.77.205415}{\bibinfo{year}{2008}}).

\bibitem{Block11}
\bibinfo{author}{T.~Block}, \bibinfo{author}{C.~Tegenkamp},
  \bibinfo{author}{J.~Baringhaus}, \bibinfo{author}{H.~Pfn\"ur}, and
  \bibinfo{author}{T.~Inaoka}, \bibinfo{title}{Plasmons in {Pb} nanowire arrays
  on {Si}(557): Between one and two dimensions},
  \bibinfo{journal}{\href{http://dx.doi.org/10.1103/PhysRevB.84.205402}{Phys.
  Rev. B}} \href{http://dx.doi.org/10.1103/PhysRevB.84.205402}{{\bf
  \bibinfo{volume}{84}}, \bibinfo{pages}{205402}}
  (\href{http://dx.doi.org/10.1103/PhysRevB.84.205402}{\bibinfo{year}{2011}}).

\bibitem{Krieg13}
\bibinfo{author}{U.~Krieg}, \bibinfo{author}{C.~Brand},
  \bibinfo{author}{C.~Tegenkamp}, and \bibinfo{author}{H.~Pfn{\"{u}}r},
  \bibinfo{title}{{One-dimensional collective excitations in Ag atomic wires
  grown on Si(557)}},
  \bibinfo{journal}{\href{http://dx.doi.org/10.1088/0953-8984/25/1/014013}{J.
  Phys.: Condens. Matter}}
  \href{http://dx.doi.org/10.1088/0953-8984/25/1/014013}{{\bf
  \bibinfo{volume}{25}}, \bibinfo{pages}{014013}}
  (\href{http://dx.doi.org/10.1088/0953-8984/25/1/014013}{\bibinfo{year}{2013}}).

\bibitem{Nagao06}
\bibinfo{author}{T.~Nagao}, \bibinfo{author}{S.~Yaginuma},
  \bibinfo{author}{T.~Inaoka}, and \bibinfo{author}{T.~Sakurai},
  \bibinfo{title}{{One-Dimensional Plasmon in an Atomic-Scale Metal Wire}},
  \bibinfo{journal}{\href{http://dx.doi.org/10.1103/PhysRevLett.97.116802}{Phys.
  Rev. Lett.}} \href{http://dx.doi.org/10.1103/PhysRevLett.97.116802}{{\bf
  \bibinfo{volume}{97}}, \bibinfo{pages}{116802}}
  (\href{http://dx.doi.org/10.1103/PhysRevLett.97.116802}{\bibinfo{year}{2006}}).

\bibitem{Lichtenstein16a}
\bibinfo{author}{T.~Lichtenstein}, \bibinfo{author}{C.~Tegenkamp}, and
  \bibinfo{author}{H.~Pfn{\"{u}}r}, \bibinfo{title}{{Lateral electronic
  screening in quasi-one-dimensional plasmons}},
  \bibinfo{journal}{\href{http://dx.doi.org/10.1088/0953-8984/28/35/354001}{J.
  Phys. Condens. Matter}}
  \href{http://dx.doi.org/10.1088/0953-8984/28/35/354001}{{\bf
  \bibinfo{volume}{28}}, \bibinfo{pages}{354001}}
  (\href{http://dx.doi.org/10.1088/0953-8984/28/35/354001}{\bibinfo{year}{2016}}).

\bibitem{Sanna18}
\bibinfo{author}{S.~Sanna}, \bibinfo{author}{T.~Lichtenstein},
  \bibinfo{author}{Z.~Mamiyev}, \bibinfo{author}{C.~Tegenkamp}, and
  \bibinfo{author}{H.~Pfn\"ur}, \bibinfo{title}{How one-dimensional are atomic
  gold chains on a substrate?},
  \bibinfo{journal}{\href{http://dx.doi.org/10.1021/acs.jpcc.8b08600}{The
  Journal of Physical Chemistry C}}
  \href{http://dx.doi.org/10.1021/acs.jpcc.8b08600}{{\bf
  \bibinfo{volume}{122}}, \bibinfo{pages}{25580}}
  (\href{http://dx.doi.org/10.1021/acs.jpcc.8b08600}{\bibinfo{year}{2018}}).

\bibitem{Lichtenstein19}
\bibinfo{author}{T.~Lichtenstein}, \bibinfo{author}{Z.~Mamiyev},
  \bibinfo{author}{E.~Jeckelmann}, \bibinfo{author}{C.~Tegenkamp}, and
  \bibinfo{author}{H.~Pfnür}, \bibinfo{title}{Anisotropic {2D} metallicity:
  plasmons in {Ge}(100)-{Au}},
  \bibinfo{journal}{\href{http://dx.doi.org/10.1088/1361-648x/ab02c5}{J. Phys.:
  Condens. Matter}} \href{http://dx.doi.org/10.1088/1361-648x/ab02c5}{{\bf
  \bibinfo{volume}{31}}, \bibinfo{pages}{175001}}
  (\href{http://dx.doi.org/10.1088/1361-648x/ab02c5}{\bibinfo{year}{2019}}).

\bibitem{Rugeramigabo10}
\bibinfo{author}{E.~P. Rugeramigabo}, \bibinfo{author}{C.~Tegenkamp},
  \bibinfo{author}{H.~Pfn\"ur}, \bibinfo{author}{T.~Inaoka}, and
  \bibinfo{author}{T.~Nagao}, \bibinfo{title}{One-dimensional plasmons in
  ultrathin metallic silicide wires of finite width},
  \bibinfo{journal}{\href{http://dx.doi.org/10.1103/PhysRevB.81.165407}{Phys.
  Rev. B}} \href{http://dx.doi.org/10.1103/PhysRevB.81.165407}{{\bf
  \bibinfo{volume}{81}}, \bibinfo{pages}{165407}}
  (\href{http://dx.doi.org/10.1103/PhysRevB.81.165407}{\bibinfo{year}{2010}}).

\bibitem{Stern1967}
\bibinfo{author}{F.~Stern}, \bibinfo{title}{Polarizability of a two-dimensional
  electron gas},
  \bibinfo{journal}{\href{http://dx.doi.org/10.1103/PhysRevLett.18.546}{Phys.
  Rev. Lett.}} \href{http://dx.doi.org/10.1103/PhysRevLett.18.546}{{\bf
  \bibinfo{volume}{18}}, \bibinfo{pages}{546}}
  (\href{http://dx.doi.org/10.1103/PhysRevLett.18.546}{\bibinfo{year}{1967}}).

\bibitem{Williams74}
\bibinfo{author}{P.~F. Williams} and \bibinfo{author}{A.~N. Bloch},
  \bibinfo{title}{Self-consisten dielectric response of a quasi-one-dimensional
  metal at high frequencies},
  \bibinfo{journal}{\href{http://dx.doi.org/10.1103/PhysRevB.10.1097}{Phys.
  Rev. B}} \href{http://dx.doi.org/10.1103/PhysRevB.10.1097}{{\bf
  \bibinfo{volume}{10}}, \bibinfo{pages}{1097}}
  (\href{http://dx.doi.org/10.1103/PhysRevB.10.1097}{\bibinfo{year}{1974}}).

\bibitem{Gold92}
\bibinfo{author}{A.~Gold}, \bibinfo{title}{Elementary excitations in multiple
  quantum wire structures},
  \bibinfo{journal}{\href{http://dx.doi.org/10.1007/BF01320939}{Z. Phys. B -
  Condensed Matter}} \href{http://dx.doi.org/10.1007/BF01320939}{{\bf
  \bibinfo{volume}{89}}, \bibinfo{pages}{213}}
  (\href{http://dx.doi.org/10.1007/BF01320939}{\bibinfo{year}{1992}}).

\bibitem{Moudgil10}
\bibinfo{author}{R.~K. Moudgil}, \bibinfo{author}{V.~Garg}, and
  \bibinfo{author}{K.~N. Pathak}, \bibinfo{title}{{Confinement and correlation
  effects on plasmons in an atom-scale metallic wire}},
  \bibinfo{journal}{\href{http://dx.doi.org/10.1088/0953-8984/22/13/135003}{J.
  Phys.: Condens. Matter}}
  \href{http://dx.doi.org/10.1088/0953-8984/22/13/135003}{{\bf
  \bibinfo{volume}{22}}, \bibinfo{pages}{135003}}
  (\href{http://dx.doi.org/10.1088/0953-8984/22/13/135003}{\bibinfo{year}{2010}}).

\bibitem{Schulz83}
\bibinfo{author}{H.~J. Schulz}, \bibinfo{title}{Long-range {C}oulomb
  interactions in quasi-one-dimensional conductors},
  \bibinfo{journal}{\href{http://dx.doi.org/10.1088/0022-3719/16/35/010}{Journal
  of Physics C: Solid State Physics}}
  \href{http://dx.doi.org/10.1088/0022-3719/16/35/010}{{\bf
  \bibinfo{volume}{16}}, \bibinfo{pages}{6769}}
  (\href{http://dx.doi.org/10.1088/0022-3719/16/35/010}{\bibinfo{year}{1983}}).

\bibitem{Kopietz95}
\bibinfo{author}{P.~Kopietz}, \bibinfo{author}{V.~Meden}, and
  \bibinfo{author}{K.~Sch\"onhammer}, \bibinfo{title}{Anomalous scaling and
  spin-charge separation in coupled chains},
  \bibinfo{journal}{\href{http://dx.doi.org/10.1103/PhysRevLett.74.2997}{Phys.
  Rev. Lett.}} \href{http://dx.doi.org/10.1103/PhysRevLett.74.2997}{{\bf
  \bibinfo{volume}{74}}, \bibinfo{pages}{2997}}
  (\href{http://dx.doi.org/10.1103/PhysRevLett.74.2997}{\bibinfo{year}{1995}}).

\bibitem{Kopietz97}
\bibinfo{author}{P.~Kopietz}, \bibinfo{author}{V.~Meden}, and
  \bibinfo{author}{K.~Sch\"onhammer}, \bibinfo{title}{Crossover between
  {L}uttinger and {F}ermi-liquid behavior in weakly coupled metallic chains},
  \bibinfo{journal}{\href{http://dx.doi.org/10.1103/PhysRevB.56.7232}{Phys.
  Rev. B}} \href{http://dx.doi.org/10.1103/PhysRevB.56.7232}{{\bf
  \bibinfo{volume}{56}}, \bibinfo{pages}{7232}}
  (\href{http://dx.doi.org/10.1103/PhysRevB.56.7232}{\bibinfo{year}{1997}}).

\bibitem{essler05}
\bibinfo{author}{F.~Essler}, \bibinfo{author}{H.~Frahm},
  \bibinfo{author}{F.~G\"{o}hmann}, \bibinfo{author}{A.~Kl\"{u}mper}, and
  \bibinfo{author}{V.~Korepin}, {\em \bibinfo{title}{The One-Dimensional
  Hubbard Model}\/} (\bibinfo{publisher}{Cambridge University Press},
  \bibinfo{address}{Cambridge}, \bibinfo{year}{2005}).

\bibitem{scho05}
\bibinfo{author}{U.~Schollw{\"o}ck}, \bibinfo{title}{The density-matrix
  renormalization group},
  \bibinfo{journal}{\href{http://dx.doi.org/10.1103/RevModPhys.77.259}{Rev.
  Mod. Phys.}} \href{http://dx.doi.org/10.1103/RevModPhys.77.259}{{\bf
  \bibinfo{volume}{77}}, \bibinfo{pages}{259}}
  (\href{http://dx.doi.org/10.1103/RevModPhys.77.259}{\bibinfo{year}{2005}}).

\bibitem{scho11}
\bibinfo{author}{U.~Schollw{\"o}ck}, \bibinfo{title}{The density-matrix
  renormalization group in the age of matrix product states},
  \bibinfo{journal}{\href{http://dx.doi.org/10.1016/j.aop.2010.09.012}{Ann.
  Phys.}} \href{http://dx.doi.org/10.1016/j.aop.2010.09.012}{{\bf
  \bibinfo{volume}{326}}, \bibinfo{pages}{96}}
  (\href{http://dx.doi.org/10.1016/j.aop.2010.09.012}{\bibinfo{year}{2011}}).

\bibitem{jeck08a}
\bibinfo{author}{E.~Jeckelmann}, {\em \bibinfo{title}{Density-Matrix
  Renormalization Group Algorithms}\/}, {\em \bibinfo{booktitle}{Computational
  Many-Particle Physics}\/}, edited by \bibinfo{editor}{H.~Fehske},
  \bibinfo{editor}{R.~Schneider}, and \bibinfo{editor}{A.~Wei{\ss}e}, volume
  \bibinfo{volume}{739} of \bibinfo{series}{Lecture Notes in Physics},
  chapter~\bibinfo{chapter}{21}, pp. \bibinfo{pages}{597--619}
  (\bibinfo{publisher}{Springer Berlin Heidelberg}, \bibinfo{year}{2008}).

\bibitem{DasSarma96}
\bibinfo{author}{S.~{Das Sarma}} and \bibinfo{author}{E.~Hwang},
  \bibinfo{title}{{Dynamical response of a one-dimensional quantum-wire
  electron system}},
  \bibinfo{journal}{\href{http://dx.doi.org/10.1103/PhysRevB.54.1936}{Phys.
  Rev. B}} \href{http://dx.doi.org/10.1103/PhysRevB.54.1936}{{\bf
  \bibinfo{volume}{54}}, \bibinfo{pages}{1936}}
  (\href{http://dx.doi.org/10.1103/PhysRevB.54.1936}{\bibinfo{year}{1996}}).

\bibitem{Wang01}
\bibinfo{author}{D.~W. Wang} and \bibinfo{author}{S.~Das~Sarma},
  \bibinfo{title}{Elementary electronic excitations in one-dimensional
  continuum and lattice systems},
  \bibinfo{journal}{\href{http://dx.doi.org/10.1103/PhysRevB.65.035103}{Phys.
  Rev. B}} \href{http://dx.doi.org/10.1103/PhysRevB.65.035103}{{\bf
  \bibinfo{volume}{65}}, \bibinfo{pages}{035103}}
  (\href{http://dx.doi.org/10.1103/PhysRevB.65.035103}{\bibinfo{year}{2001}}).

\bibitem{Chou2020}
\bibinfo{author}{Y.-Z. Chou} and \bibinfo{author}{S.~Das~Sarma},
  \bibinfo{title}{Nonmonotonic plasmon dispersion in strongly interacting
  {C}oulomb {L}uttinger liquids},
  \bibinfo{journal}{\href{http://dx.doi.org/10.1103/PhysRevB.101.075430}{Phys.
  Rev. B}} \href{http://dx.doi.org/10.1103/PhysRevB.101.075430}{{\bf
  \bibinfo{volume}{101}}, \bibinfo{pages}{075430}}
  (\href{http://dx.doi.org/10.1103/PhysRevB.101.075430}{\bibinfo{year}{2020}}).

\bibitem{bent07}
\bibinfo{author}{H.~Benthien} and \bibinfo{author}{E.~Jeckelmann},
  \bibinfo{title}{Spin and charge dynamics of the one-dimensional extended
  {H}ubbard model}, \bibinfo{journal}{Phys. Rev. B} {\bf \bibinfo{volume}{75}},
  \bibinfo{eid}{205128}  (\bibinfo{year}{2007}).

\bibitem{jeck02a}
\bibinfo{author}{E.~Jeckelmann}, \bibinfo{title}{Dynamical density-matrix
  renormaliza\-tion-group method},
  \bibinfo{journal}{\href{http://dx.doi.org/10.1103/PhysRevB.66.045114}{Phys.
  Rev. B}} \href{http://dx.doi.org/10.1103/PhysRevB.66.045114}{{\bf
  \bibinfo{volume}{66}}, \bibinfo{pages}{045114}}
  (\href{http://dx.doi.org/10.1103/PhysRevB.66.045114}{\bibinfo{year}{2002}}).

\bibitem{Mila1993}
\bibinfo{author}{F.~Mila} and \bibinfo{author}{X.~Zotos}, \bibinfo{title}{Phase
  diagram of the one-dimensional extended hubbard model at quarter-filling},
  \bibinfo{journal}{\href{http://dx.doi.org/10.1209/0295-5075/24/2/010}{Europhys.
  Lett.}} \href{http://dx.doi.org/10.1209/0295-5075/24/2/010}{{\bf
  \bibinfo{volume}{24}}, \bibinfo{pages}{133}}
  (\href{http://dx.doi.org/10.1209/0295-5075/24/2/010}{\bibinfo{year}{1993}}).

\bibitem{Penc1994}
\bibinfo{author}{K.~Penc} and \bibinfo{author}{F.~Mila}, \bibinfo{title}{Phase
  diagram of the one-dimensional extended {H}ubbard model with attractive
  and/or repulsive interactions at quarter filling},
  \bibinfo{journal}{\href{http://dx.doi.org/10.1103/PhysRevB.49.9670}{Phys.
  Rev. B}} \href{http://dx.doi.org/10.1103/PhysRevB.49.9670}{{\bf
  \bibinfo{volume}{49}}, \bibinfo{pages}{9670}}
  (\href{http://dx.doi.org/10.1103/PhysRevB.49.9670}{\bibinfo{year}{1994}}).

\bibitem{Ejima2005}
\bibinfo{author}{S.~Ejima}, \bibinfo{author}{F.~Gebhard}, and
  \bibinfo{author}{S.~Nishimoto}, \bibinfo{title}{Tomonaga-{L}uttinger
  parameters for doped {M}ott insulators},
  \bibinfo{journal}{\href{http://dx.doi.org/10.1209/epl/i2005-10020-8}{Europhys.
  Lett.}} \href{http://dx.doi.org/10.1209/epl/i2005-10020-8}{{\bf
  \bibinfo{volume}{70}}, \bibinfo{pages}{492}}
  (\href{http://dx.doi.org/10.1209/epl/i2005-10020-8}{\bibinfo{year}{2005}}).

\bibitem{Shirakawa2009}
\bibinfo{author}{T.~Shirakawa} and \bibinfo{author}{E.~Jeckelmann},
  \bibinfo{title}{Charge and spin {D}rude weight of the one-dimensional
  extended {H}ubbard model at quarter filling},
  \bibinfo{journal}{\href{http://dx.doi.org/10.1103/PhysRevB.79.195121}{Phys.
  Rev. B}} \href{http://dx.doi.org/10.1103/PhysRevB.79.195121}{{\bf
  \bibinfo{volume}{79}}, \bibinfo{pages}{195121}}
  (\href{http://dx.doi.org/10.1103/PhysRevB.79.195121}{\bibinfo{year}{2009}}).

\bibitem{Meden2000}
\bibinfo{author}{V.~Meden}, \bibinfo{author}{W.~Metzner},
  \bibinfo{author}{U.~Schollw{\"o}ck}, \bibinfo{author}{O.~Schneider},
  \bibinfo{author}{T.~Stauber}, and \bibinfo{author}{K.~Sch{\"o}nhammer},
  \bibinfo{title}{Luttinger liquids with boundaries: {P}ower-laws and energy
  scales},
  \bibinfo{journal}{\href{http://dx.doi.org/10.1007/s100510070180}{Eur. Phys.
  J. B}} \href{http://dx.doi.org/10.1007/s100510070180}{{\bf
  \bibinfo{volume}{16}}, \bibinfo{pages}{631}}
  (\href{http://dx.doi.org/10.1007/s100510070180}{\bibinfo{year}{2000}}).

\bibitem{Andergassen2006}
\bibinfo{author}{S.~Andergassen}, \bibinfo{author}{T.~Enss},
  \bibinfo{author}{V.~Meden}, \bibinfo{author}{W.~Metzner},
  \bibinfo{author}{U.~Schollw{\"o}ck}, and
  \bibinfo{author}{K.~Sch{\"o}nhammer}, \bibinfo{title}{Renormalization-group
  analysis of the one-dimensional extended {H}ubbard model with a single
  impurity},
  \bibinfo{journal}{\href{http://dx.doi.org/10.1103/PhysRevB.73.045125}{Phys.
  Rev. B}} \href{http://dx.doi.org/10.1103/PhysRevB.73.045125}{{\bf
  \bibinfo{volume}{73}}, \bibinfo{pages}{045125}}
  (\href{http://dx.doi.org/10.1103/PhysRevB.73.045125}{\bibinfo{year}{2006}}).

\bibitem{jeck13}
\bibinfo{author}{E.~Jeckelmann}, \bibinfo{title}{Local density of states of the
  one-dimensional spinless fermion model},
  \bibinfo{journal}{\href{http://dx.doi.org/10.1088/0953-8984/25/1/014002}{J.
  Phys.: Condens. Matter}}
  \href{http://dx.doi.org/10.1088/0953-8984/25/1/014002}{{\bf
  \bibinfo{volume}{25}}, \bibinfo{pages}{014002}}
  (\href{http://dx.doi.org/10.1088/0953-8984/25/1/014002}{\bibinfo{year}{2013}}).

\bibitem{Abdelwahab18a}
\bibinfo{author}{A.~Abdelwahab} and \bibinfo{author}{E.~Jeckelmann},
  \bibinfo{title}{Luttinger liquid and charge density wave phases in a spinless
  fermion wire on a semiconducting substrate},
  \bibinfo{journal}{\href{http://dx.doi.org/10.1103/PhysRevB.98.235138}{Phys.
  Rev. B}} \href{http://dx.doi.org/10.1103/PhysRevB.98.235138}{{\bf
  \bibinfo{volume}{98}}, \bibinfo{pages}{235138}}
  (\href{http://dx.doi.org/10.1103/PhysRevB.98.235138}{\bibinfo{year}{2018}}).

\end{thebibliography}

\end{document}